\documentclass[conference]{IEEEtran}
\IEEEoverridecommandlockouts
\usepackage{cite}
\usepackage{amsmath,amssymb,amsfonts}
\usepackage{algorithmic}
\usepackage{graphicx}
\usepackage{textcomp}
\usepackage{xcolor}
\def\BibTeX{{\rm B\kern-.05em{\sc i\kern-.025em b}\kern-.08em
    T\kern-.1667em\lower.7ex\hbox{E}\kern-.125emX}}

\usepackage{pgfpages}
\pgfpagesuselayout{resize to}[a4paper]   
    
\begin{document}

\title{Proactive Tasks Management based on a Deep Learning Model
}

\author{\IEEEauthorblockN{Kostas Kolomvatsos}
\IEEEauthorblockA{\textit{Department of Informatics and Telecommunications} \\
\textit{University of Thessaly}\\
Papasiopoulou 2-4, 35131 Lamia Greece \\
kostasks@uth.gr}
\and
\IEEEauthorblockN{Christos Anagnostopoulos}
\IEEEauthorblockA{\textit{School of Computing Science} \\
\textit{University of Glasgow}\\
Glasgow, UK \\
christos.anagnostopoulos@glasgow.ac.uk}

}

\maketitle

\begin{abstract}
Pervasive computing applications deal with intelligence surrounding users 
that can facilitate their activities. 
This intelligence is provided in the form of software components 
incorporated in embedded systems or devices 
in close distance with end users.
One example infrastructure that can host intelligent pervasive services 
is the Edge Computing (EC) infrastructure. 
EC nodes can execute a number of tasks for data collected by devices present 
in the Internet of Things (IoT) infrastructure. 
In this paper, we propose 
an intelligent, proactive tasks management model based on the demand.
Demand depicts the number of users or applications interested in 
using the available tasks in EC nodes, thus, characterizing their 
popularity. 
We rely on a Deep Machine Learning (DML) model and more specifically 
on a Long Short Term Memory (LSTM) network to 
learn the distribution of demand indicators for each task and estimate the
future interest. 
This information is combined with historical observations and support
a decision making scheme to conclude which tasks will be offloaded due to limited 
interest on them. We have to notice that in our decision making, we also take 
into consideration the load that every task may add
to the processing node where it will be allocated.
The description of our model is accompanied by 
a large set of experimental simulations for evaluating the proposed mechanism.
We provide numerical results and reveal that the proposed scheme is capable of deciding on the fly while concluding the most efficient allocation.
\end{abstract}


\section{Introduction}
The advent of the Internet of Things (IoT) offers many opportunities to 
the development of novel applications over a huge infrastructure of numerous devices.
These devices are directly connected with 
the Edge Computing (EC) ecosystem to report the collected data and 
consume the provided services.
At the EC, one can meet nodes with processing capabilities that assist in the
provision of services with the minimum possible 
latency to end users.
The main reason for that is that processing activities
are kept close to end users.
In addition, processing at the edge can reduce the network traffic \cite{shi} driving data analytics towards geo-distributed processing, known as edge analytics \cite{pu}, \cite{Satyanarayanan}, \cite{yi}, \cite{zhou}. 
The discussed activities take the form of a set of tasks
that should be executed by EC nodes. 
However, EC nodes are characterized by heterogeneity in their computational resources and, more importantly,
by different load. The dynamic environment of the IoT makes the load for 
EC nodes fluctuating not only in terms of numbers but also in terms of 
the computational burden that a task may add to an EC node.
Past efforts in the field \cite{kolomvatsos1} deal with models
that can be adopted to deliver the computational complexity of 
tasks, thus, we can have an estimate on the burden that tasks may cause to processing nodes.

Due to the dynamic nature where the ecosystem of EC nodes and IoT devices act, the requirements of tasks,
their number and the demand for them are continuously changing.
For alleviating EC nodes from an increased load towards delivering the final response in the minimum possible time, EC nodes may decide to offload a sub-set of tasks to their peers or Cloud.
Efficient solutions should be provided that will allocate tasks to the available processing nodes to conclude the desired processing activities (e.g., the provision of analytics)
\cite{Kathidjiotis}.
This way, we can easily support real time applications keeping the quality of service for end users at high levels. 
Tasks offloading was first appeared in Mobile Cloud Computing (MCC) \cite{fernando}, \cite{sanaei}.
MCC targets to offload tasks from mobile nodes to Cloud data centers where centralized
computing and storage take place.  
Some obstacles for this model are related to the 
delay in sending tasks and getting responses especially when communications are realized 
over a Wide Area Network (WAN). 
Moreover, the variability of the contextual information of EC nodes, tasks and the collected data define strict requirements for 
the effective conclusion of the allocation of tasks. 
Another significant obstacle is related to 
the heterogeneity of the EC nodes. 
Any offloading action should be realized upon the 
dynamic nodes' contextual information, thus, 
any proposed scheme should meet all the imposed requirements.
In any case, the decision for offloading tasks and the 
finalization of the specific allocations should be the result
of a monitoring process and a reasoning action locally at the EC nodes.
The challenge is to select the appropriate tasks to be offloaded to peers or Cloud.
Any decision should be mandated by the 
following targets: (i) maximize the performance and (ii) minimize the consumption of resources.
Finally, any decision should be realized in a distributed manner, i.e., EC nodes independently 
decide their line of actions.
Multiple research efforts deal with centralized approaches, however, these allocation and scheduling models suffer from the drawbacks reported in the literature for Cloud computing \cite{islam}. 

In this paper, we focus on the problem of offloading tasks to peer nodes or Cloud and go beyond the state of the art compared to our previous effort in the domain \cite{karanika}. 
Instead of using an uncertainty management methodology (in \cite{karanika}, we propose a model built upon the principles of Fuzzy Logic), we investigate the adoption of 
Deep Machine Learning (DML) technologies
\cite{goodfellow}. Our current orientation is to focus on a completely different 
technique for solving the aforementioned problem.
More specifically, we build upon a Long Short Term Memory (LSTM) Recurrent Neural Network (RNN) 
and expose a decision making mechanism for selecting the appropriate tasks to be offloaded 
by a EC node. We combine the outcome of the LSTM model with a scheme based on the 
multi-criteria decision making theory \cite{franco}.
Our `reasoning' is dictated by the demand that end users or IoT devices exhibit for every task.
It is a strategic decision for our model to incorporate the `popularity' of tasks
in the selection process towards supporting popular tasks to be kept locally instead 
of being offloaded to other nodes.
The intuition behind this is two fold: First, nodes save resources through the re-use of the tasks execution framework; Secondly, the latency experienced by users is minimized as highly demanded tasks are initiated and executed immediately. 
We consider that EC nodes 
record 
the demand for each task as being affected by the 
mobility of end users/IoT devices.
It becomes obvious that the discussed mobility 
opens up the road for imposing spatio-temporal requirements in our model. 
Additionally, the mobility of end users 
increases the complexity of the reasoning mechanisms when trying to find out if a task will be kept locally and adds uncertainty in nodes' behaviour. 
The proposed approach is also characterized by the necessary 
scalability as it can efficiently support an increased number of users.
The reason is that we support EC nodes with a pre-trained DML model, thus,
EC nodes can easily apply the envision reasoning process no matter the number of
end users.  
We also have to notice that an EC node may decide to keep the execution of a task locally no matter its popularity when the load it will add is very low. This aspect is incorporated into our rewarding mechanism that affects the ranking of each task
before an offloading action is decided. 
The following list depicts the contributions of our paper: 
\begin{itemize}
 \item We propose a georeferenced task management scheme where computation offloading is decided based on tasks demand;
 \item We adopt a DML, i.e., LSTM model to estimate the future demand for each task present in an EC node;
 \item We provide an `aggregation' mechanism that combines past demand observations and future estimates to feed our reasoning mechanism and decide the tasks that should be offloaded to peers/Cloud;
 \item We support the `reasoning' mechanism of EC nodes adopting the principles of the multi-criteria theory;
 \item We provide an extensive experimental evaluation that reveals the pros and cons of the proposed approach. Our evaluation is performed for a set of metrics adopting real traces. 
\end{itemize}
Results indicate that our model is capable of supporting real time applications while exhibiting an increased performance for a large set of experimental scenarios. 

The remaining paper is organized as follows.
Section \ref{section2} reports on the related work and presents 
important research efforts in the field.
In Section \ref{section3}, we discuss preliminary information and describe our problem
while in Section \ref{section4}, we present the proposed mechanism. Section \ref{section5} is 
devoted to the description of our experimental evaluation adopting 
a set of performance metrics.
Finally, in Section \ref{section6}, we conclude our paper giving our future research plans. 

\section{Related Work}
\label{section2}
The advent of the EC comes into scene to offer a `cover' of the 
IoT infrastructure giving the opportunity of adopting an additional
processing layer before the collected data be transferred to Cloud.
Numerous EC nodes can create an ecosystem 
of autonomous devices capable of interacting with IoT devices and
themselves to execute a set of tasks.
Tasks are processing activities requested by 
by applications or end users
and can be of any form. 
For instance, tasks can request for the delivery of Machine Learning (ML) models
(e.g., regression, clustering) 
or ask for the execution of `typical' SQL-like queries over the available data.
The advantage of EC is that these processing activities can be realized 
close to end users,
thus, limiting the latency they enjoy
\cite{lin}.
One can say that it is the best way to keep the processing at the EC ecosystem 
as long as possible before relying on Cloud. 
We have to create a cooperative ecosystem that 
makes EC nodes capable of interacting to execute the requested tasks.
This cooperation may involve the offloading of tasks to peers.
The most significant reasons for that are the high load that an EC node may face, the 
absence of the necessary computational capabilities or the lack of the appropriate data.

Research community is very active in the field of tasks management in a large set of 
applications domains.
Recent studies deal with 
tasks offloading solutions, i.e., partitioning, allocation, 
resource management and distributed execution \cite{lin}.
The offloading action belongs to one of the following modes:
(i) full offloading and (ii) partial offloading \cite{wang2}.
In the former mode, tasks will be executed as a whole 
no matter the location avoiding to partition each task. 
For instance, we could adopt a model that 
delivers the appropriate place to offload 
the desired tasks based on various characteristics (tasks and 
nodes) \cite{sardellitti}. 
The latter mode builds on the parallel execution of 
a set of sub-tasks (a partitioning process is adopted for that) possibly 
offloaded in different processing nodes.
Additional efforts deal with joint tasks allocation,
i.e., the allocation of tasks requested by different users/devices/applications
\cite{dab}. 
The target is to minimize the trade off between the performance when executing 
tasks and meeting the constraints of nodes (e.g., energy resources \cite{zhou}).
This means that we try to gain from executing tasks requested by multiple 
users/applications which can be considered as a type of resource sharing \cite{shi}.
An example of a resource sharing model is presented in 
\cite{kao} where a polynomial-time task assignment scheme is proposed 
for allocating tasks with inter-dependency towards achieving 
guaranteed latency-energy trade offs. 

ML is also adopted in a set efforts dealing with tasks offloading.
Reinforcement learning is a candidate solution that can 
lead to the best possible action upon a rewarding mechanism \cite{dab}.
Tasks allocation can be also studied as an optimization problem
\cite{du} where constraints can depict the 
monetary or time costs for solving the problem \cite{wang}.
The discussed problem can be formulated as a maximization (maximize the reward)
or a minimization (minimize the cost for every allocation) process.
In any case, a high number of constraints make the optimization approach 
an NP-hard problem `dictating' the adoption of an approximate solution
or the use of a set of assumptions.

Various schemes have been proposed for supporting the efficient tasks
allocation. In \cite{dong}, a dynamic, decentralized
resource-allocation strategy based on evolutionary game theory is presented.
The matching theory is adopted in \cite{gu}, i.e.,
the model does not take into consideration the central Cloud in the Mobile Edge Computing (MEC) 
platform
considering the autonomous nature of edge nodes
\cite{anagnostopoulos1}. 
A coalition-game-based cooperative method to optimize the problem of task offloading is the subject of \cite{zhang} while in \cite{josilo}, the authors
present game-based strategies
for the discussed problem to achieve the Nash equilibrium among mobile users. 
In \cite{sheng}, the authors discuss a model 
for computation offloading under a scenario
of multi-user and multi-mobile edge servers that considers the performance of intelligent devices
and server resources. 
The task scheduling part of the model is based on an auction algorithm
by considering the time requirements of the computing tasks and the performance
of the mobile edge server. 
In 
\cite{xing}, the authors propose a device-to-device (D2D)-
enabled multi-helper MEC system, 
in which a local user offloads its tasks to multiple helpers
for cooperative computation. 
The model tries to minimize the latency by
optimizing the local user's task assignment jointly with the time and rate for task offloading and results
downloading, as well as the computation frequency for task execution.
In any case, the proposed approaches should take into consideration the 
characteristics of the dynamic environment where EC nodes and IoT devices act.
In \cite{zhang2}, the authors focus on  
an access control management architecture for a 5G heterogeneous network.
Two algorithms are considered, i.e., an optimal static algorithm based on
dynamic programming and 
a two-stage online algorithm to adaptively
obtain the current optimal solution in real time. 
Dynamic programming is also adopted in \cite{wu}
while integer linear programming is proposed 
in \cite{cuervo}. A randomized version of the dynamic programming approach 
is proposed in  
\cite{shahzad}.  

Additional research efforts deal with the `cooperation' of 
EC, IoT and Cloud \cite{ning}.
In such a setting, the offloading action can be performed taking into consideration 
multiple layers as we can meet multiple points where processing
activities can be realized.
Some example technologies adopted to solve the problems 
focusing on multiple `entities' 
are  
the branch and bound algorithm for delivering approximate solutions, 
the Mixed Integer Linear Programming (MILP),
the Iterative Heuristic MEC
Resource Allocation (IHRA) algorithm and so on and so forth.
All of them result dynamic decisions based on the realization of every parameter.
In \cite{kim}, the authors consider 
the estimation of the total processing time
of each task and for each candidate processing node using linear
regression. The same approach is adopted 
in the effort presented in 
\cite{messaoudi}.
It is critical to estimate the 
time requirements of tasks 
(e.g., the round trip time - RTT) taking into consideration 
all the necessary nodes' and network's parameters.
If we have in our hands the estimated 
time, we can easily deliver the 
burden that every task will cause to a processing node, thus, we can 
decide fully aligned with the real needs.

Recent developments deal with the advent of 
Software Defined Networking (SDN) and the need of coordinating 
virtualized resources \cite{misra}.
The optimality of the decision is related 
to the local or remote task computation, the selection of the 
appropriate node and the selection of the appropriate path for 
the offloading action. 
In \cite{lin2}, the authors study the flexible compute-intensive task offloading to a local Cloud
trying to optimize energy consumption, operation speed, and cost. 
In \cite{alghandi}, a model based on the Optimal Stopping Theory (OST) is adopted to 
deliver the appropriate time to offload data and tasks to an
edge server. The challenge is to to determine the best offloading strategy
that minimizes the expected total delay. 
Finally, in \cite{callegaro}, the authors consider
unmanned vehicles (i.e., Unmanned Aerial Vehicles - UAVs) and 
propose a framework enabling optimal offloading decisions as a function of 
network \& computation load as well as the current state. 
The optimization is formulated as an optimal stopping time problem over a Markov process.

\section{Preliminaries \& Problem Formulation}
\label{section3}
For the description of the problem under consideration, we borrow the notation provided 
in \cite{karanika} that deals with the same problem, however, it proposes a different solution compared to the current effort.
We consider  
a set of $N$ EC nodes, $\mathcal{N} = \left\lbrace n_{1}, n_{2}, \ldots, n_{N} \right\rbrace$,
`connected' with a number of IoT devices 
being responsible to collect and store data while performing
the requested tasks.
EC nodes stand in the middle between the IoT devices and Cloud undertaking the responsibility of 
receiving the reported data transferring them upwards to the Cloud for further processing.
EC nodes become the host of geo-distributed datasets
giving the opportunity of performing the execution of processing activities close to end users. 
Such processing activities may be requested by end users or applications 
and deal with the execution of tasks either simple or complex.
For instance, an application may ask for analytics related to the collected data in a spatiotemporal 
basis to realize additional services for users.
We have to notice that EC nodes, compared to Cloud, are characterized by 
limited computational resources, thus, the execution of tasks should be carefully decided.
Concerning the collected data, EC nodes should also host the necessary software for storing, processing and retrieving them.
It becomes obvious that every EC node can be transformed to a `intelligent' entity that 
takes decisions related to the management of data and tasks on the fly.  

Tasks execution aims at generating knowledge locally.
As tasks may belong to different types (e.g., simple SQL-like queries, ML models generation, etc),
they impose different processing requirements for EC nodes. 
In our past research efforts \cite{kolomvatsos1}, we provide
a specific methodology for exposing 
the computational burden that a task may impose to EC nodes.
Without loss of generality, we consider that nodes may support the 
same number of tasks, i.e., $E$. 
At a time instance $t$, an EC node may have to execute a subset of the 
aforementioned tasks. For their management, we consider that EC nodes 
adopt a queue where tasks are placed just after their arrival.
A node retrieves the first task from the queue and proceeds with its execution.
The number of tasks executed in a time unit defines the 
throughput of EC nodes and depends on their computational capabilities, thus,
it consequently
affects the size of the queue.
Tasks present in the queue define the future load of each node being adopted 
(as we will see later) to estimate if it is feasible
to service all the waiting tasks in a reasonable time interval.
A task may be requested by a number of users/applications
indicating the need for repeated executions of it.
When a task is popular, EC nodes may re-use the pre-executed instances 
and limit the time for delivering a response while they save resources.
This is significant when tasks share the same parameters 
(e.g., multiple tasks request for the same regression analysis over the hosted data)
and the same requests.
Additionally, tasks may be characterized by a specific priority
(e.g., the same approach as adopted in real time 
operating systems for serving processes) especially in the case when 
real time applications should receive the final response as soon as possible.
However, the management of priorities is a complex issue as some tasks 
(especially those with a low priority) may suffer from the starvation effect.
In this effort we do not take into account 
a `preemptive' scheme, i.e., a task with a high priority 
may interrupt the execution of tasks with a low priority.
This approach is considered in the first place of our future research plans.

\begin{figure}[h]
\centering
\includegraphics[width=0.5\textwidth]{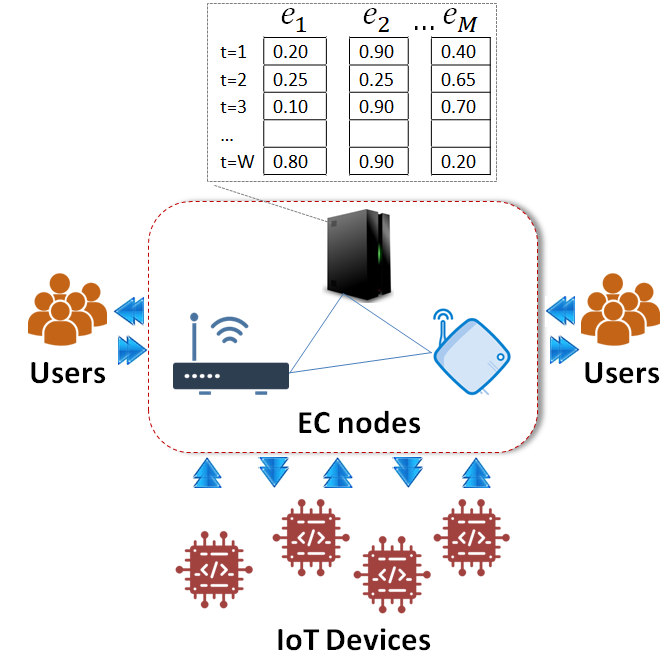}
\caption{An example of the proposed architecture}
\label{fig1}
\end{figure}

As mentioned above, EC nodes are capable of estimating the future load 
based on tasks present in the queue.
In case the future load cannot be efficiently served, EC nodes may decide 
to offload some of the already present tasks.
This is a critical decision as it affects their future performance as well as 
the time for delivering the final response to users/applications.
In this paper, we propose to take into consideration the demand for each task before we decide 
those that will be offloaded to peers/Cloud.
The rationale is to keep the processing locally for tasks requested by many users/applications
to re-use the pre-executed activities as explained above.
Hence, there will be more room to release resources for the execution
of the remaining tasks before additional requests arrive.
Let us consider that the demand for a task is represented by a value
in the unity interval, i.e., a value close to unity depicts a high demand 
and the opposite stands for a value close to zero.
We focus on the behaviour of a single EC node, i.e., $n_{i}$ (the same approach holds true for every node in the EC infrastructure). 
At $t$, $n_{i}$ observes the demand for each task and stores it 
in a dedicated data structure (e.g., a vector).
Let this vector be the \textit{Tasks Demand Vector} (TDV),
$TDV = \left\lbrace e_{1}^{t}, e_{2}^{t}, \ldots, e_{M}^{t} \right\rbrace$ 
with $M \leq E$.
We have to notice that $n_{i}$ may not process the entire set of the $E$ available 
tasks in the network but only a sub-set. In any case, this observation does not affect our 
model.
Hence, for every monitoring epoch, i.e.,
$t = 1, 2, \ldots$, $n_{i}$ updates the corresponding TDV
and maintains the $W$ latest recordings (see Figure \ref{fig1}).
This is a sliding window approach as $n_{i}$ wants to 
keep only `fresh' information about the demand of every task.
Consider that the estimated future load for tasks present 
in the queue (the calculation of the future estimated load can be performed 
as indicated in \cite{kolomvatsos1}) indicates that their execution is beyond the current 
`capabilities' of $n_{i}$.
In this case, $n_{i}$ should decide to `evict' a number of the $M$ available tasks 
to peers/Cloud. The final allocation of the evicted tasks
can be performed as described in our previous efforts
\cite{kolomvatsos}.

We provide a solution to the critical research question of which tasks should be selected
to be offloaded to peers/Cloud.
We aim to keep the execution locally for popular tasks in order to eliminate more
the time for providing the responses.
By offloading non popular tasks, nodes may save resources as they are 
not benefited from re-using previous outcomes. Additionally,
EC nodes may accept a slightly increased latency for non-popular 
tasks to release more resources for popular tasks.
The rationale behind this strategic orientation of our model is simple.
A non-popular task may be offloaded to another node that may have 
increased demand for it (incremental models and caching may be adopted 
to deliver the final result) paying only the communication cost (for sending the task and getting 
the response) and the time for waiting the final outcome.
In any case, our model should be accompanied by the appropriate scheme for selecting 
the right peer for sending the task as proposed in \cite{kolomvatsos}.
Our focus is to rely on the 
available TDVs and the load that every task causes in $n_{i}$ 
before we perform the final selection. 
For this, we propose the adoption of historical demand observations and the combination of a DML 
with the principles of multi-criteria decision making theory.

\section{Tasks Management at the Edge}
\label{section4}

\subsection{Tasks Demand Indicator}
EC nodes rely on TDVs to decide whether a task should be offloaded
to peers.
$n_{i}$ applies a monitoring scheme for 
updating the local TDVs,
i.e.,
$\left\lbrace TDV^{t}\right\rbrace$ with
$TDV^{t} = \left\lbrace e_{1}^{t}, e_{2}^{t}, \ldots, e_{M}^{t} \right\rbrace$ 
and $t \in \left\lbrace 1, 2, \ldots, W \right\rbrace$.
At specific epochs, $n_{i}$ updates the $\left\lbrace TDV^{t}\right\rbrace$
by evicting the oldest observations and storing the most recent ones.
Based on $\left\lbrace TDV^{t}\right\rbrace_{t=1}^{W}$,
$n_{i}$ can estimate the future demand for each task
and define the \textit{Demand Indicator} (DI).
As we will see later, the DI affects the 
final ranking of each task as delivered by a function $f(\cdot)$.
$f(\cdot)$ realizes the DI of a task as exposed by past observations and future estimates.
Formally, the DI could be defined as the number of 
users/applications requesting a specific task,
i.e., $DI_{j} = e_{j}^{n_{i}}$.
We argue that the DI can be discerned as the DI exposed by past observations $DI_{p}$ and the DI
expose by an estimation process $DI_{f}$. Both, $DI_{p}$ \& $DI_{f}$ are aggregated to deliver the 
final DI for each task, i.e., $DI_{F}$.
The $DI_{p}$ is delivered by a function $g(\cdot)$ that gets the last demand values
(e.g., three) 
$e_{j}^{t}, t \in \left\lbrace W-l \right\rbrace, l=0,1,\ldots$ for the $j$th task.
$g(\cdot)$ can be any function we desire to 
clearly depict the recent demand observations.
The outcome of the discussed function, i.e., $DI_{p} = g\left( e_{j}^{t} \right) \in \mathbb{R}^{+}, t \in \left\lbrace W-k \right\rbrace$ is aggregated in a subsequent step with $DI_{f}$ to deliver the
$DI_{F}$. Then, $DI_{F}$ values are adopted 
in the proposed multi-criteria rewarding mechanism to result the 
\textit{Offloading Degree} (OD) of a task.
$OD_{j}, j =1, 2, \ldots, M$ are fed
into the function $f(\cdot)$ which is a ranking function to result the
final sorted list of the available tasks.
In our model, the last-$k$ tasks in the aforementioned list are selected to be offloaded 
in peer nodes as proposed in \cite{kolomvatsos}.

\subsection{The Proposed LSTM}
Based on the observed TDVs, $n_{i}$ is able to 
process a `time series' dataset for the $j$th task,
i.e., $e_{j}^{1}, e_{j}^{2}, \ldots, e_{j}^{W}$.
Upon this sequence of demand values, we are able to estimate future values e.g., $e_{j}^{W+1}, e_{j}^{W+2}, \ldots$.
Our aim is to combine what is experienced so far (exposed by TDVs) with the expected 
realizations of demand.
We select to adopt an LSTM \cite{goodfellow}, i.e., a specific type of RNNs to capture the demand trends for each task.
Our LSTM tries to `understand' every demand value based on previous realizations and efficiently learn the distribution of data.
Legacy neural networks cannot perform well in cases where we want to capture the trend of a time series.
RNNs and LSTMs are network with loops inside of them making data to persist. 
We have to notice that the LSTM delivers $DI_{f}$ for each task being present in the queue of $n_{i}$.
In our model, we adopt an LSTM for the following reasons:
(i) we want to give the opportunity to the proposed model to learn over large sequences  of data ($W >> 1$) and not only over recent data. Typical RNNs suffer from short-term memory and may leave significant information from the beginning of the sequence making difficult the transfer of information from early steps to the later ones;
(ii) typical RNNs also suffer from the vanishing gradient problem, i.e., when a gradient becomes very low during back propagation, the network stops to learn;
(iii) LSTMs perform better the processing of data compared to other architectures as they incorporate multiple `gates' adopted to regulate the flow of the information. Hence, they can learn better than other models upon time series.

Every LSTM cell in the architecture of the network has an internal recurrence (i.e., a self-loop) in addition to the external recurrence of typical RNNs.
It also has more parameters than an RNN 
and the aforementioned gates to control the flow of data.
The self-loop weight is controller by the so-called forget gate, i.e.,
$g_{f}^{t} = \sigma \left( b^{f} + \sum_{j} U_{j}^{f} \mathbf{e}_{j}^{t} + \sum_{j} Z_{j}^{f} h_{j}^{t-1}\right)$   
where 
$\sigma$ is the standard deviation of the unit, $b^{f}$ represents the bias of the unit,
$U^{f}$ represents the input weights,
$\mathbf{e}$ is the vector of inputs
(we can get as many inputs as we want out of $W$ recordings),
$Z^{f}$ represents the weights of the forget gate and $h^{t-1}$ represents the current hidden layer vector.
The internal state of an LSTM cell is updated as follows:
$s^{t} = g_{f}^{t} s^{t-1} + g_{in}^{t} \sigma \left( b + \sum_{j} U_{j} \mathbf{e}_{j}^{t} + \sum_{j} Z_{j} h_{j}^{t-1}\right)$.
Now, 
$b$, $U$ and $Z$ represent the bias,
input weights and recurrent weights of the cell
and $g_{in}$ depicts the external input gate.
We perform similar calculations for the external input $g_{in}$ and the output gates
$g_{out}$. 
The following equations hold true:
\begin{equation}
g_{in}^{t} = \sigma \left( b^{in} + \sum_{j} U_{j}^{in} \mathbf{e}_{j}^{t} + \sum_{j} Z_{j}^{in} h_{j}^{t-1}\right)
\end{equation}

\begin{equation}
g_{out}^{t} = \sigma \left( b^{out} + \sum_{j} U_{j}^{out} \mathbf{e}_{j}^{t} + \sum_{j} Z_{j}^{out} h_{j}^{t-1}\right)
\end{equation}

The output of the cell is calculated as follows:

\begin{equation}
h^{t} = tanh \left( s^{t} \right) g_{out}^{t}
\end{equation}

We adopt a multiple input, single output LSTM where the final output represents the 
estimated demand value at $W+1$, i.e.,
$e_{j}^{W+1}$ for the $j$th task.
$DI_{f} = e_{j}^{W+1}$ is, then, combined with the 
past observations to deliver an efficient decision making mechanism for selecting the tasks that will be offloaded to peers.
We have to notice that the LSTM model is trained upon real datasets as we discuss in the experimental evaluation section.

\subsection{Aggregating Past Observations \& Future Estimates}
Having calculated $DI_{p}$ and $DI_{f}$,
we result the final $DI_{F}$ for each task
in the queue.
$DI_{F}$ is the outcome 
of a function 
$c(\cdot)$ that gets $DI_{p}$ and $DI_{f}$ and delivers a value 
in the unity interval,
i.e., $DI = c \left( DI_{p}, DI_{f} \right) \to [0,1]$.
When $DI_{F} \to 1$ means that 
the specific task exhibits a high demand as exposed by past observations and future estimates.
The opposite stands for the scenario where 
$DI_{F} \to 0$.
The aggregation of $DI_{p}$ and $DI_{f}$ is performed 
through the adoption of the 
Weighted Geometric Mean (WGM) \cite{Jung}. 
The WGM is calculated as follows:
\begin{equation}
\label{WGM}
    DI_{F}= e^{\left(\frac{\sum_{i=1}^{|l|} w_{i} ln(DI_{i})}{\sum_{i=1}^{|l|} w_{i}} \right)} 
\end{equation}
with $l = \left\lbrace p, f \right\rbrace$ and 
In Eq(\ref{WGM}), $w_{i} \in [0,1], \sum w_{i} = 1$ represents weight of each demand indicator.
$w_{i}$ is selected to depict the strategy we want to adopt in the delivery of the outcomes.
For instance, 
if $w_{i=p} \to 1$, the model pays more attention 
on the past observations instead of relying on the LSTM result.
The opposite stands for $w_{i=f} \to 1$.

\subsection{The Proposed Rewarding Scheme \& Decision Making}
We adopt a rewarding 
mechanism for extracting the reward that $n_{i}$ gains if it 
executes locally a task. 
Tasks with the lowest reward will be offloaded to peer nodes.
We rely on multiple rewards, 
i.e., $R = \left\lbrace r_{j} \right\rbrace \in \mathbb{R}^{+}, \mathcal{EN}|, j = 1, 2, \ldots$; one for each 
parameter affecting the final decision.
In our model, we consider two rewards for $DI_{F}$ and for the load $\lambda$ that 
every task will add in the hosting node.
For $DI_{F}$, we consider that when $DI_{F} \geq T_{DI}$, we gain a reward $r_{1}$;
otherwise we pay a penalty equal to $r_{1}$.
$T_{DI}$ is a pre-defined threshold for comparing 
the $DI_{F}$ value.
Through the specific approach, we aim at keeping locally the execution of 
popular tasks as already explained.
The same approach stands also for $\lambda$. 
When $\lambda \leq T_{\lambda}$, the specific tasks gets a reward equal to 
$r_{2}$; otherwise, the task gets a penalty equal to $r_{2}$.
$T_{\lambda}$ is the pre-defined threshold that indicates 
when a reward/penalty should be assigned to a task.
We have to notice that the reward/penalty for $\lambda$ is considered only when 
the queue size is over a specific percentage assuming that the maximum queue
size is equal to $Q_{max}$. 
In general, the proposed methodology is adopted only when 
$n_{i}$ faces an increased load and it has to offload some tasks to avoid its overloading that will negatively affect the performance (in a dynamic environment, more tasks will continue to arrive).
For both rewards, we apply a sigmoid function for `smoothing' the outcome,
i.e., $r^{s}_{j} = r_{j} \cdot \sum \frac{1}{1+e^{-(\gamma y - \delta)}}$ where $\gamma$ and $\delta$ are parameters adopted 
to `calibrate' its shape.
In the above equation, $y$ represents the difference between the aforementioned pairs
of parameters with the corresponding thresholds, 
i.e., $y \in \left\lbrace DI_{F} - T_{DI}, T_{\lambda} - \lambda  \right\rbrace$.
The higher the difference is, the higher the reward becomes.
For instance, when $DI_{F} >> T_{DI}$, the corresponding reward may quickly approach 
unity or zero in the scenario where $DI_{F} << T_{DI}$.
The same rationale stands for the $\lambda$ parameter compared to its 
threshold $T_{\lambda}$.
The final reward for the $j$th task, i.e., the aforementioned OD, is calculated as follows:
$OD = r^{final}_{j} = \sum_{\forall j} r_{j}$.
Tasks present in the $n_{i}$'s queue are sorted by 
$r^{final}$ and the last-$k$ task are offloaded to peers.

\section{Experimental Evaluation}
\label{section5}

\subsection{Performance Indicators \& Setup}
We report on the performance of the proposed model as far as the 
conclusion of correct offloading decision concerns.
We also focus on time requirements to result the final decision for tasks that will be offloaded to peer nodes.
We plan to expose the ability of the proposed scheme to get real time
decisions to be able to support time critical applications.
We perform a high number of experiments and get the mean value for a set of 
performance metrics.
We evaluate the proposed model upon the simulation of the demand realizations
based on a real trace.
For simulating the demand for each task, we rely on the 
dataset discussed in \cite{tsanas}.
The dataset is generated by an energy analysis 
of 12 different building shapes, i.e.,
their glazing area, glazing area distribution and their orientation. 
From this dataset, we `borrow' the data related to the temperature load of 
each building to represent the demand for our tasks.
We have to notice that the dataset has an `attitude' to low 
values. 

We adopt a set of performance metrics for evaluating our model.
These metrics are as follows:
\textbf{(i)} the average time $\tau$ spent to conclude a decision. $\tau$ is measured for every task as the time spent (CPU time) by the system deciding if a task should be offloaded or not. For this reason, $\tau$ is calculated as the sum of (a) the time spent to get the outcome of the LSTM; (b) the time spent for the rewarding mechanism to deliver the final reward for each task; (c) the time required to deliver the final rankings of tasks and select those that will be offloaded to peers. We have to notice that 
$\tau$ is measured as the mean required time per task in seconds; 
\textbf{(ii)} the number of correct decisions $\Delta$. For realizing $\Delta$,
we assume the cost for executing a tasks locally compared to the cost 
for offloading the task. The cost for executing locally a task is equal to the waiting time in the queue plus the execution time.
The cost for offloading a tasks involves the migration cost, the waiting time in the `remote' queue and the time required for getting the response from the peer node.
It becomes obvious that depending on the performance of the network the `typical' case is to have a higher cost when offloading a task. However, EC nodes can undertake this cost for non popular tasks if it is to release resources for assisting in the execution of popular tasks.
Hence, we consider a correct decision as the decision that offloads tasks that their $DI_{F}$ is below the pre-defined threshold $T_{DI}$, i.e., 
\begin{equation}
\Delta = \frac{\sum \frac{|DI_{F} < T_{DI}|}{k}}{EX}
\end{equation} 
where $EX$ is the number of experiments. Recall that at every epoch we offload the last-$k$ tasks of the ranked list. Hence, $\Delta$ depicts the percentage of $k$ tasks that are correctly offloaded based on the reasoning of our decision making mechanism;
\textbf{(iii)} we adopt the $\omega$ metric that depicts the percentage of the offloaded tasks that are among the $k$ tasks with the smallest popularity. We try to figure out if the proposed model can detect non popular tasks. We have to remind that, in our decision making, the demand/popularity is combined with the load that a task adds to an EC node. We strategically decide to keep the execution of non popular tasks locally when the load they add is very low.

We perform a set of experiments for different $W$, $E$, $w_{i=p}$ and $T_{DI}$. We adopt $W\in \left\lbrace 50, 100 \right\rbrace$, i.e., different sliding window sizes to measure the effect on $\tau$, $\Delta$ and $\omega$. The total number of tasks requested by the users is set to $E \in \left\lbrace 500, 1000, 5000\right\rbrace$. Moreover, the weight of past observations is adopted 
as $w_{i=p} \in \left\lbrace 0.3, 0.7 \right\rbrace$. The probability of having the demand for a task over a pre-defined threshold is set to $T_{DI} = 0.5$. In total, we conduct 100 iterations for each experiment and report our results for the aforementioned metrics. The experiments are executed using Python and an Intel i7 CPU with 16Gb Ram.

\subsection{Performance Assessment}
\begin{figure}[h]
\centering
\includegraphics[width=0.5\textwidth]{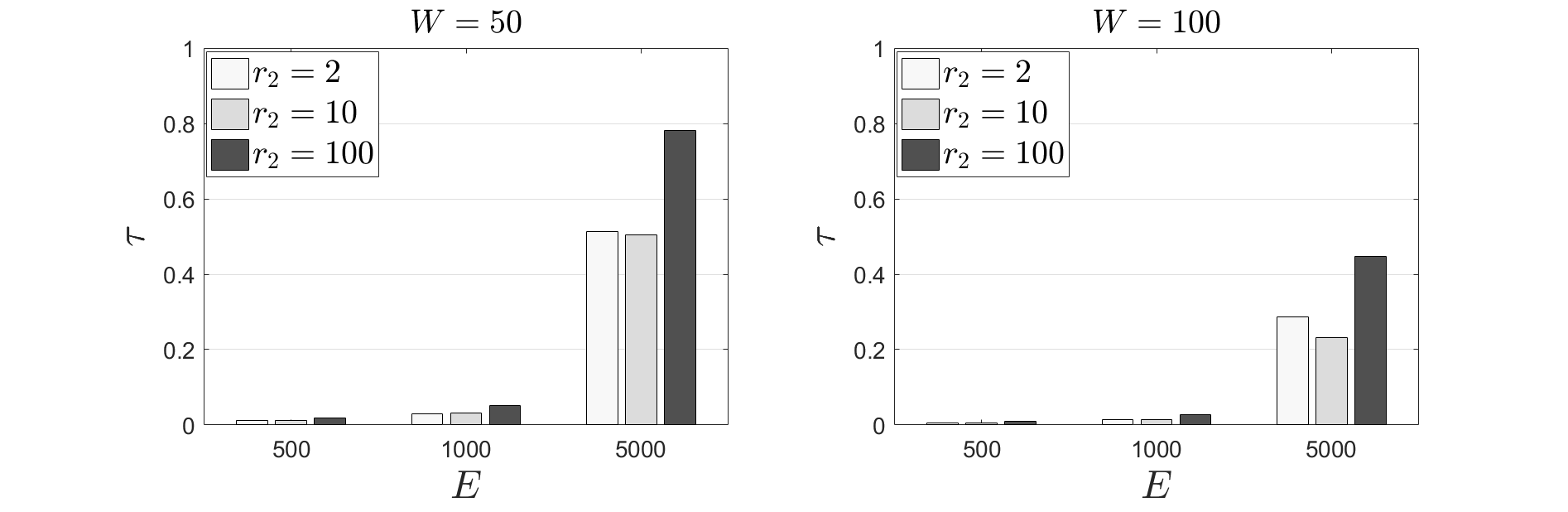}
\caption{Performance outcomes for the $\tau$ metric when $k=3$}
\label{fig2}
\end{figure}

\begin{figure}[h]
\centering
\includegraphics[width=0.5\textwidth]{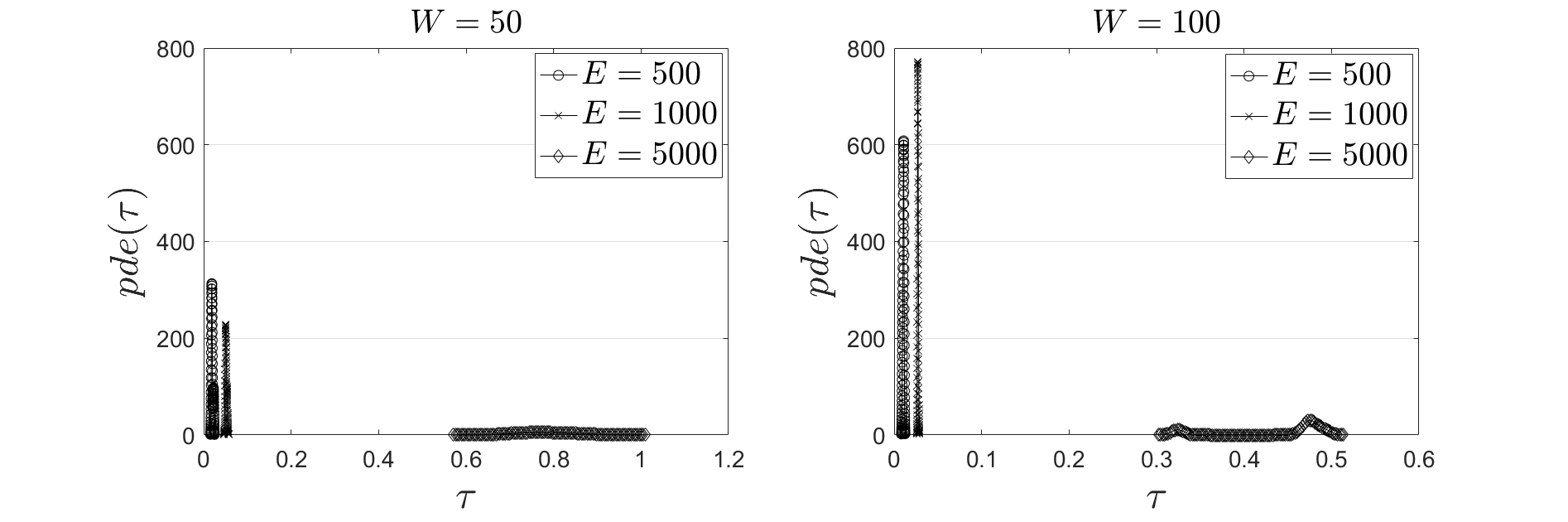}
\caption{The pde of $\tau$ for various experimental scenarios}
\label{fig2-1}
\end{figure}

We report on the performance of the proposed model related to the $\tau$ metric. In this set of experiments, we keep $k=3$, i.e., every EC node should `evict' only a small sub-set of the available tasks and $w_{i=p} = 0.7$. 
In Figure \ref{fig2}, we present our outcomes for 
$W \in \left\lbrace 50, 100 \right\rbrace$ and for different rewards for the load of each task
($r_{2} \in \left\lbrace 2, 10, 100 \right\rbrace$).
We observe that $E$ (the number of tasks) heavily affects the outcome of $\tau$. An increased number of tasks leads to an increased mean conclusion time per task. Additionally, 
the size of the window is inversely proportional to the mean required time, i.e., a low $W$ leads to an increased
$\tau$ and vice versa. 
These results are naturally extracted; an EC node has to process too many tasks, it requires more time to perform the 
calculations mandated by our model. 
To elaborate more on the performance evaluation for 
the $\tau$ metric, in Figure \ref{fig2-1},
we present the probability density estimate (pde)
of the required time to conclude the final decision.
We actually confirm our previous observations.
The proposed model requires around a second (in average) to process 5000 tasks when the sliding window is small.
In the case of a large window, our scheme requires 
0.4 seconds (in average) to process 5000 tasks.
The remaining evaluation outcomes reveal that when
$E < 5000$ at each EC node,
it is possible to manage the requested tasks in time below 0.1 seconds (in average).
This exhibits the ability of the proposed model 
to react in serving the needs of real time applications requesting the execution of tasks in high rates.
This is because, we can pre-train the proposed LSTM scheme, then, upload it at the available EC nodes to be adopted to conclude the offloading decisions.
We have to notice that the training process lasts for
(for 1000 epochs) around 2.5 minutes. 
Obviously, the training process can be realized in EC nodes with an increased frequency (if necessary) without jeopardizing their functioning. 
It should be also noticed that $\Delta$ is equal to unity for all the experimental scenarios no matter the values of the adopted parameters. This means that the demand of the selected tasks to be offloaded in peer nodes is below the pre-defined threshold $T_{DI}$, thus, no popular tasks 
are evicted.
Recall that the final decision also takes into consideration the load that every task causes into the hosting node and nodes are eager to keep locally tasks with a very low load.  

\begin{figure}[h]
\centering
\includegraphics[width=0.5\textwidth]{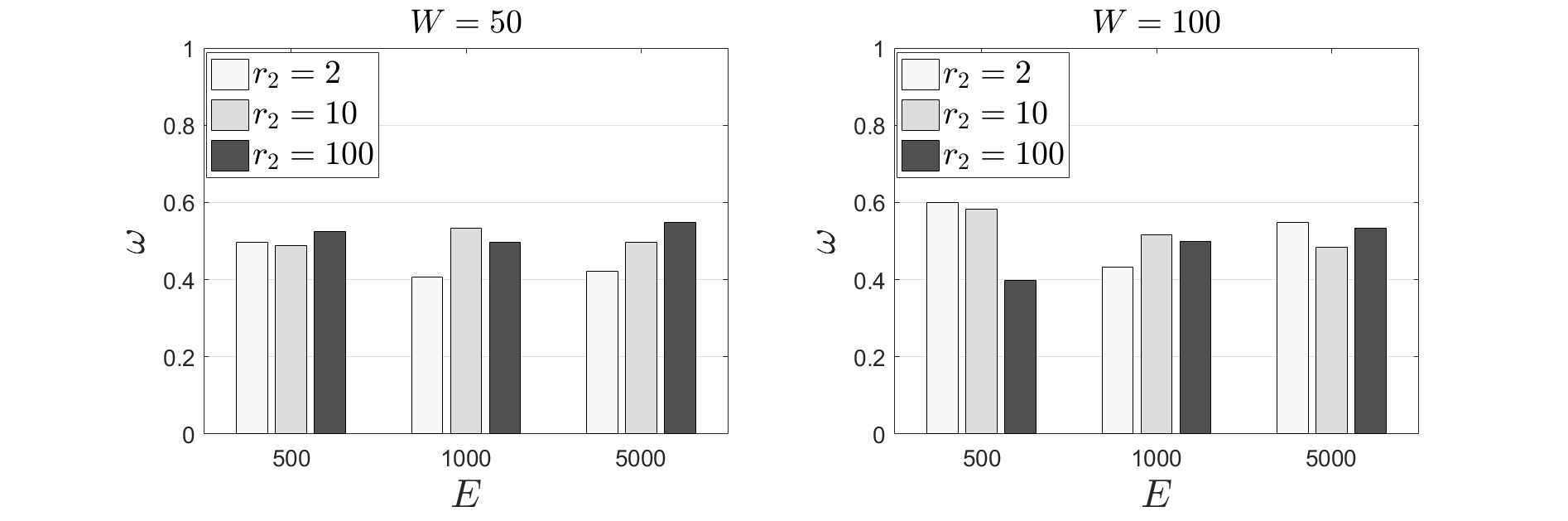}
\caption{Performance outcomes for the $\omega$ metric when $k=3$}
\label{fig3}
\end{figure}

In Figure \ref{fig3}, we present our results for the $\omega$ metric.
We observe that half (approximately) of the selected tasks 
are between those with the minimum popularity. Our outcomes are similar no matter the size of the sliding window. The same stands true when we focus on the number of tasks.
It seems that all the aforementioned parameters are not heavily affecting the final selection.
All the retrieved results are in the interval [0.4, 0.6].
We also conclude that the effect of $\lambda$ and the corresponding reward does not let $\omega$ to get high values. For instance, a task may have a low popularity, however, a very low load as well. 
This task will get an increased reward and will not be in the last-$k$ tasks that will be offloaded
to peers.

\begin{figure}[h]
\centering
\includegraphics[width=0.5\textwidth]{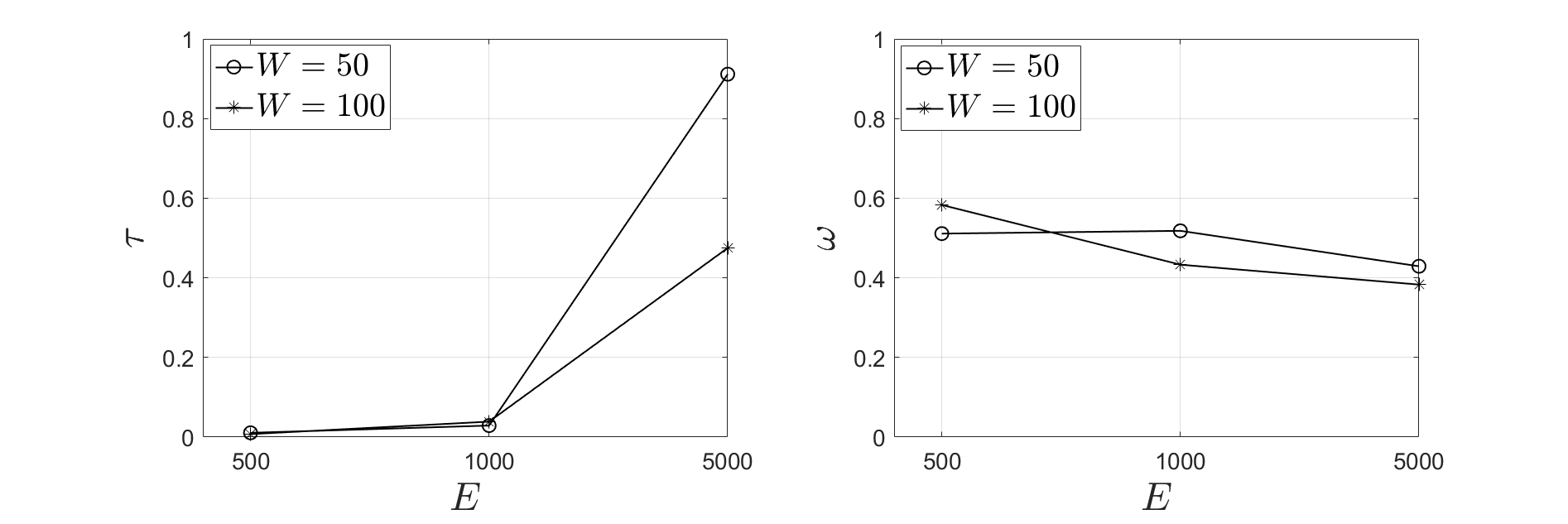}
\caption{Performance outcomes for $\tau$ and $\omega$ metrics when $k=3$ and $w_{i=p}=0.3$}
\label{fig4}
\end{figure}

We perform a set of experiments adopting $w_{i=p} = 0.3$. Now, the focus of our decision making mechanism is on the demand estimation retrieved by the proposed LSTM model. In Figure \ref{fig4}, we present our results for 
the $\tau$ and $\omega$ metrics.
Again, we observe an increased conclusion time when 
$E \to 5000$.
$\omega$ decreases as $E \to 5000$ exhibiting more clearly the effect of the incorporation of $\lambda$ in the 
rewarding scheme, thus, in the decision making model. 
The best results are achieved for a limited sliding window size, i.e., $W=50$.
We have to also notice that $\Delta$ is equal to unity as in the previous set of experiments.

\begin{figure}[h]
\centering
\includegraphics[width=0.5\textwidth]{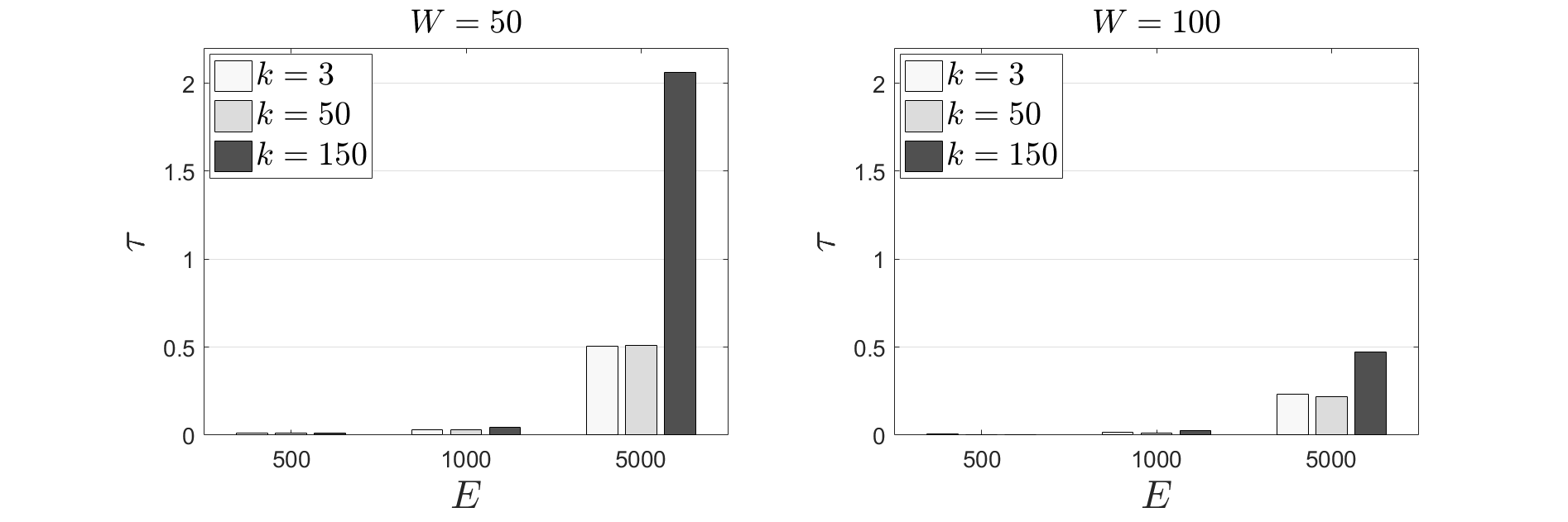}
\caption{The time requirements of the proposed model for different $k$ values}
\label{fig5}
\end{figure}

\begin{figure}[h]
\centering
\includegraphics[width=0.5\textwidth]{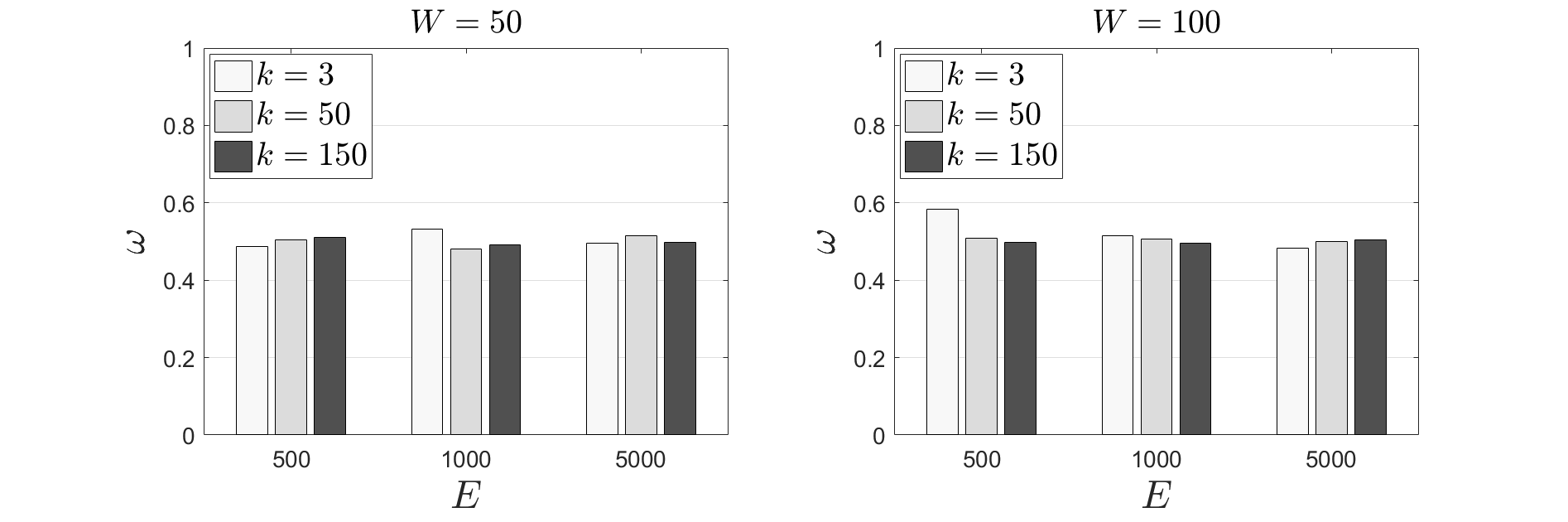}
\caption{The performance evaluation of the proposed model for the $\omega$ metric and different $k$ values}
\label{fig6}
\end{figure}

In the following experimental scenarios, we adopt different $k$ values, i.e., different number of 
tasks that should be offloaded to peers.
Figure \ref{fig5} depicts our results.
Naturally, increased $k$ and $E$ negatively affect the time requirements of our mechanism. The worst case scenario is met when $k=150$ and $E=500$. 
In this case, the proposed mechanism needs around 2 seconds per task (in average) to extract the final decision.
Concerning the $\omega$ metric (Figure \ref{fig6}),
we get similar results as in the previous experimental scenarios. 
Again, around half of the available tasks selected to be offloaded are among those that exhibit the lowest popularity.
Finally, $\Delta$ realization is equal to unity except the scenario where $E=500$ and $k=150$.
In this scenario, EC nodes have to evict the 30\% (approximately) of the available tasks.
In such cases, $\Delta$ is equal to 0.69 and 0.89
for $W \in \left\lbrace 50, 100\right\rbrace$, respectively.  
Now, some tasks with demand over $T_{DI}$ may be offloaded;
a decision that is affected by the incorporation of $\lambda$ in the rewarding scheme.
Evidently, some tasks with low popularity, however, with a low load as well, may be kept locally.
This decision is fully aligned with the above discussed strategic design of the behaviour of EC nodes.

We compare the performance of our model
with the scheme presented in \cite{baranidharan} where
the authors 
propose a task scheduling algorithm (ETSI)
that is based on a heuristic.
This heuristic delivers the final outcome
based on the remaining energy, the distance 
from the edge of the network and the 
number of neighbours calculating the rank of each node.
The node with the lowest ranking 
is selected for the final allocation.
We additionally compare our scheme with the model
presented in \cite{karanika}.
There, a fuzzy logic model is adopted to decide which tasks
will be offloaded to peers.
The comparison is performed for the $\Delta$ metric.
Our scheme outperforms both models. 
For instance, ETSI manages to result a limited number of
correct decisions related to the offloading of tasks.
The highest realization of $\Delta$
is 45\% (approximately) with the mean and median be 
around 25\%.
Moreover, the lowest value for 
$\Delta$ in \cite{karanika}  
is around 80\% depending on the experimental scenario.
The proposed model exhibits worse performance than the scheme in \cite{karanika} only when EC nodes should evict 
too many tasks (like in the scenario when $k=150$, $E=500$ and $W=50$). 

\section{Conclusions \& Future Work}
\label{section6}
Tasks scheduling and offloading actions are significant for a number of application domains.
The performance of tasks' execution may be enhanced if we rely on a cooperative model that makes processing nodes to interact and exchange tasks. 
This way, nodes can release resources to serve the remaining tasks reported by users or applications.
In this paper, we focus on the discussed problem and take into consideration the dynamic nature of environments like the IoT or the EC where nodes interact. 
We focus on the behaviour of EC nodes related to the management of tasks.
We support their behaviour with an intelligent scheme that decides which tasks should be offloaded to peer nodes.
We incorporate into the proposed model a deep learning scheme and a rewarding mechanism. Both technologies aim to detect tasks that should be kept locally based on the demand that users/applications exhibit for them.
We propose this strategy to benefit from the re-use of the resources and build upon an incremental processing approach towards the minimization of time in the provision of the final responses.
We perform an extensive set of simulations and reveal the ability of the proposed scheme to be adopted in real time setups while being aligned with the dynamics of the environment.
We present numerical results that exhibit a limited time for training the deep learning model and concluding the final list of tasks that should offloaded to peer nodes.
In the first place of our future research agenda is to apply an optimal stopping model for selecting the evicted tasks combined with the outcomes of the deep learning approach. This way, we will be able to create a more robust mechanism that will incorporate the necessary stochastic behaviour in the decision making process. 


\end{document}